\documentclass[aps,prl,twocolumn,showpacs]{revtex4}

\usepackage[dvips]{graphicx}
\usepackage{psfrag}
\usepackage{amsmath}

\begin{document}

\title{Antiferromagnetic Ordering on CoO(001) Studied by\\
Metastable Helium Beam Diffraction}
\author{P. Banerjee}
\author{X. Liu}
\author{M. Farzaneh}
\author{C.R. Willis}
\author{W. Franzen}
\author{M. El-Batanouny}
\affiliation{Department of Physics, Boston University, 590
  Commonwealth Avenue, Boston MA 02215}
\author{V. Staemmler}
\affiliation{Lehrstuhl f\"ur Theoretische Chemie,
Ruhr-Universit\"at Bochum, Germany}

\begin{abstract} 
Magnetic diffractive scattering of coherent metastable helium
atomic beams from CoO(001) surfaces reveal the presence of an
anomalous enhancement in the intensity of the half-order
antiferromagnetic diffraction peak over the temperature range of
250-320K. Electronic cluster calculations based on Hartree-Fock
states with configuration interactions, together with mean-field
and Monte Carlo studies of this system suggest that this behavior
is associated with low-lying surface spin excited states.
\end{abstract}
\maketitle

The magnetic properties of surfaces and thin films are currently a
topic of much interest.  This arises from the fact that these
properties may be completely different from their bulk counterparts.
For example, a pure surface magnetic ordering transition can occur at
a temperature T$_c^s$ higher than the bulk transition temperature
T$_c^b$. A possible occurrence of such a phenomenon in gadolinium and
terbium has been the subject of several
papers\cite{Weller,Tang,Rau88}. More recently, our group reported on
the critical behavior of antiferromagnetic (AFM) ordering on the
NiO(001) surface\cite{MM99}, as measured by metastable He beam
(He$^{\ast}$) diffraction.  These measurements have shown a surface
N\'eel temperature higher than the bulk value.  Further, the
temperature dependence of the sublattice magnetization showed a
cross-over behavior consistent with the {\sl extraordinary surface
transition} belonging to the universality class of the semi-infinite
anisotropic Heisenberg model\cite{Diehl84}.

In this Letter, we show that the AFM ordering on the CoO(001)
surface, as measured by He$^*$ diffraction, exhibits an anomalous
enhancement over the temperature range between 250 K and 320 K,
straddling the bulk N\'eel temperature T$_N^b$ of 290 K.  At
T$_N^b$, the enhancement is suppressed and a dip in the magnetic
diffraction peak intensity is observed.

\begin{figure}[tb]
\begin{center}
\psfrag{x1}[c]{$\Delta k$ (\AA$^{-1}$)}
\psfrag{x2}[c]{Intensity (arb. units)}
\includegraphics*[width=2.75in]{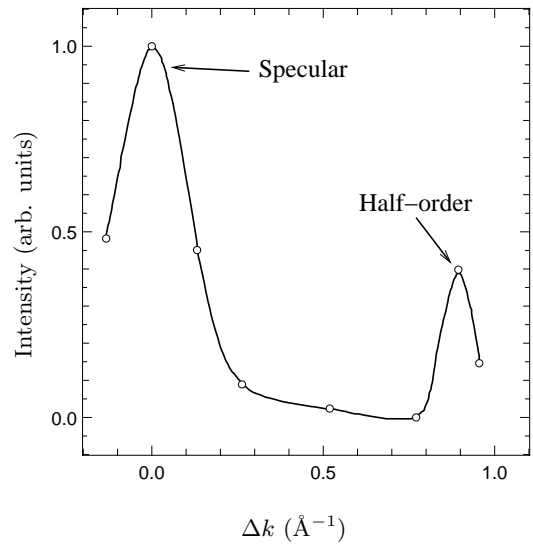}
\end{center}
\caption{\label{half}Diffraction spectra measured along the
$\langle01\rangle$ direction on CoO(001).}
\end{figure}

CoO belongs to the family of AFM 3d transition-metal monoxides
with a rocksalt structure in the paramagnetic phase.  The
spin-ordering in the bulk is characterized by ferromagnetic (111)
spin sheets that are stacked antiferromagnetically\cite{Roth58_1}.
On the (001) surface, this arrangement leads to a (2$\times$1) AFM
spin structure. Our experimental He$^*$ diffraction measurements, shown in
Fig.~\ref{half}, display half-order magnetic peaks along the
$\langle10\rangle$ and $\langle01\rangle$ directions, and no 
($1\over 2$,$1\over 2$) peak along the $\langle11\rangle$ direction, not
shown, and support the existence of such an AFM structure on the CoO(001)
surface.

It has been established \cite{Wang95, Kanamori} that the lowest bulk
crystal-field (CF) split orbital state, $^4{\bf\Gamma}_4\ (^4T_{1g})$,
has an effective angular momentum $l=1$ and is twelve-fold degenerate.
This degeneracy gives rise to a strong Jahn-Teller instability, which
is stabilized via spin-orbit (SO) interactions, giving rise to a
$j=1/2$ ground-state (GS), and $j$ = 3/2, 5/2 excited-state (ES)
manifolds.  In the GS, a hole occupies a $Y_2^{\pm1}$-type orbital,
which reduces the repulsion between a Co$^{2+}$ ion and its
neighboring O$^{2-}$ ion.  This leads to an appreciable tetragonal
contraction of $\sim1.2\%$(at low temperatures)along one of the
principal crystallographic axes\cite{Kanamori}; about two
orders of magnitude greater than the magnetostriction observed in the
other monoxides\cite{Phillips67}. Recently we have carried out
electronic cluster calculations employing Hartree-Fock states with
configuration interactions for bulk CoO which confirm all the above
features. A detailed description of the methodology and results are
given in Ref.~\cite{Staemmler}.

In earlier publications\cite{MM99,Swan93,Swan94,MM95,elbat02} we
demonstrated that He$^*$ diffraction is very sensitive to surface AFM
ordering.  In particular, in Ref.~\cite{MM99} we have shown that the
microscopic origin of He$^*$ magnetic scattering can be accounted for
by an imaginary, spin-dependent component of the He$^*$- surface
interaction potential of the form $ \Delta V^S({\bf r})\ \propto\ i\
{\cal D}\left[S({\bf R});z\right]\ \left<S_z({\bf R},T)\right>$, where
$\left<S_z({\bf R},T)\right>$ is the thermally-averaged $z$-component
of the spin at $({\bf R},z)$ and ${\cal D}\left[S({\bf R});z\right]$
is the corresponding local surface density of states. Within the
framework of the eikonal approximation, the diffractive scattering
amplitude $\tilde{\cal A}_{\bf G}$, resulting from the above imaginary
potential, together with a corrugated hard-wall real component of the
surface potential, with a corrugation shape function $\zeta({\bf R})$,
is given by
\begin{equation}
\begin{split} 
\tilde{\cal A}_{\bf G} = {e^{-(W+\alpha)}\over\Omega}
\int_{U.C.}\!\! d{\bf R}
\exp\big\{i\big[&{\bf G}\cdot{\bf R} + q_z\zeta({\bf
R})\big]  \\
& \quad -\xi({\bf R})\ \langle\hat S_z({\bf R})\rangle\big\}\ 
\end{split}
\end{equation}
where $\Omega$ is the area of the surface primitive mesh, $W$ is
the Debye-Waller factor and $\alpha$ is an attenuation factor
accounting for He$^*$ beam decay through the Penning ionization
channel.  $\xi({\bf R})$ includes ${\cal D}\left[S({\bf
R});z\right]$ plus other constants. In the case of the
(2$\times$1) spin-ordering, the diffraction amplitude of the
$1\over2$-order peak is related to the average sublattice
magnetization by \cite{MM99}
\begin{equation} 
{\cal A}_{(1/2,0)} = e^{-(W+\alpha)}
\ I_1(x),
\label{subl-mag}
\end{equation} 
where $x\propto\langle\hat S_z({\bf R})\rangle$ and $I_1(x)$ is a
modified Bessel function of order one.

The details of the design and operation of our experimental facility
have been presented earlier in Ref.~\cite{MM99}.  Beam intensities of
about $3\times10^{5}$ He$^*$ atoms s$^{-1}$ at the sample surface were
achieved in the present measurements.  All of the data were obtained
from CoO(001) surfaces freshly cleaved in vacuum, with a background
pressure less than $10^{-10}$ Torr with He beam turned off.  The CoO
sample rods were attached to a copper sample holder with silver
conducting epoxy.  A silicon diode was attached at the base of the
sample holder to measure and control the sample temperature.  The
temperature controller (Scientific Instruments Inc. Model 9600-1) is
factory calibrated and is accurate to $\pm 0.5$ K.  Scans were taken
at 2 K increments between 250 K and 320 K and at larger increments at
lower temperatures.  Measurements of the diffraction intensities
consistently yielded a minimum in the $1\over2$-order magnetic
diffraction peak intensity at T$_N^b\simeq$ 290K for all cleaved
surfaces, irrespective of their distance from the silicon diode.  We
determined T$_N^b$ to be $290\pm0.2$ K by measuring the specific heat
of several bulk samples as a function of temperature with a
Perkin-Elmer model DSC7 differential scanning calorimeter.

\begin{figure}[tb]
\begin{center}
\psfrag{x1}{T$_N^b$}
\includegraphics*[width=2.75in]{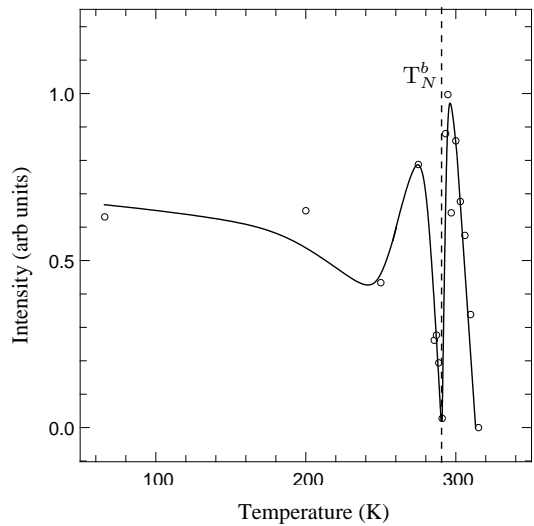}
\end{center}
\caption{\label{ho1}Half-order peak intensity as a function of surface
temperature.  Data shown is from a single cleaved surface. The dashed
line demarks the bulk N\'eel temperature.  The solid line is a guide
to the eye.}
\end{figure}

\begin{figure}[tb]
\begin{center}
\psfrag{x1}[c]{Temperature (K)}
\psfrag{x2}[c]{Intensity (arb. units)}
\psfrag{x3}{T$_N^b$}
\includegraphics*[width=2.75in]{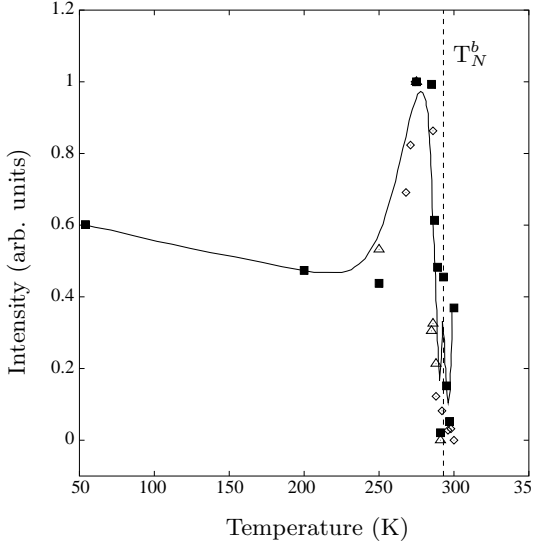}
\end{center}
\caption{\label{ho2}Intensity of the magnetic half-order diffraction
peak as a function of surface temperature.  Data shown is collected
from several different cleaved surfaces, as indicated by solid and
open circles and triangles.The dashed line demarks the bulk N\'eel
temperature.  The solid line is a guide to the eye.}
\end{figure}

Fig.~\ref{ho1} represents data collected from a single cleave and
displays the enhancement of the peak intensity both below and above
T$_N^b$. Fig.~\ref{ho2} depicts data collected from several cleaves
and shows more details of the enhancement below T$_N^b$. We notice
that the peak intensity decreases slightly with increasing temperature
from 50 K to about 250 K, but then exhibits an anomalous increase with
further increase in temperature, reaching a maximum at about 280 K,
followed by a steep decrease, reaching a minimum at T$_N^b$.  Above
T$_N^b$ it increases once more to a maximum at 310 K and finally
disappears at about 320 K.  This behavior has been checked repeatedly
and was always reproducible even for different crystal ingots.

To investigate these anomalous effects, first, we extended our
electronic cluster calculations to the CoO(001) surface.  Under the
$C_{4v}$ symmetry of the surface, the $^4{\bf\Gamma}_4$ state splits
into a nondegenerate $A_2$ and doubly-degenerate $E$ orbitals,
separated by 50 meV, comparable to SO interaction energies, so that
the latter cannot be treated as a perturbation on the CF state. A
balanced calculation shows that all the ensuing states are Kramers
doublets. The first ES doublets lie at 29 and 94 meV above the GS. We
also find that the expectation value of the spin in the GS and first
ES to be $\langle s_z\rangle \simeq 1/2$ and $\langle s_z\rangle
\simeq 3/2$, with $z$ normal to the surface.\cite{Staemmler}

Next, we constructed a model Hamiltonian that describes the low-lying
spin states and includes magnetoelastic effects. We started with the
SO eigenstates of the Co$^{+2}$ as a basis set and focused on the
lowest-lying manifold $^4F_{3/2}$. By using the operator-equivalent
formalism (OEF)\cite{Bleaney}, a spin representation for the surface
CF Hamiltonian is obtained as ${\cal H}_{CF} = c S_z^2 $.  ${\cal
H}_{CF}$ splits the $^4F_{3/2}$ manifold into two doublets:
$m_s=\pm3/2$ and $m_s=\pm1/2$ with an energy gap of $\Delta_0=2c=29$
meV. We added the effect of the tetragonal lattice distortion on
the crystal field, by introducing a uniform surface tetragonal strain
field $\eta$, which, in the OEF, has the form $b\eta S_z^2$, where $b$
is the magnetoelastic constant, and modifies ${\cal H}_{CF}$ as $
{\cal H}_{CF} = \left({\Delta_0\over2} + b\eta\right) S_z^2 $.  The
model Hamiltonian can then be written as
\begin{equation}
\begin{split}
{\cal H} =\sum_{\alpha} &{\cal H}_{CF}^\alpha + \sum_\beta {\cal
H}_{CF}^\beta  +
J_2\sum_{\stackrel{<nnn>}{\alpha\beta}}{\bf S}_{\alpha}\cdot{\bf
S}_\beta \\
& + {\kappa\over2}\ \eta^2\ +\ \sigma\eta\ ,
\end{split}
\end{equation}
where $\alpha,\beta$ denote the two AFM sublattices and $\sigma$
represents a uniform bulk tetragonal stress field that couples to
the surface strain; it has the experimentally measured
temperature-dependent form \cite{Jauch,Rechtin71} $\sigma=
\sigma_0\left(1-\frac{T}{T_N^b}\right)^{0.6} $.  We evaluated
the free energy of the system using the Bogoliubov-Peierls mean
field variational method\cite{Feynman} with a trial
noninteracting Hamiltonian of the form 
\begin{equation}
\begin{split}
{\cal H}_0 &= \sum_{\alpha} \biggl({\cal H}_{CF}^\alpha - h_\alpha
S_\alpha^z \biggr) + \sum_\beta \biggl({\cal H}_{CF}^\beta -
h_\beta\ S_\beta^z\ \biggr) \\
& \quad + {\kappa\over2}\ \eta^2\ +\ \sigma\eta\ ,
\end{split}
\end{equation}
where $h_{\alpha,\beta}$ are the effective mean fields along the
$z$-axis, taken as variational parameters.  The temperature-dependence of
the sublattice magnetization is then determined by solving the coupled
self-consistent equations of the magnetization, $M(T)$:
\begin{equation}  
M=\frac{1}{2}\,\frac{\sinh(2\beta J_2 M)+3 e^{-\beta\Delta}\sinh(6\beta
J_2 M)}{\cosh(2\beta J_2 M )+e^{-\beta\Delta}\cosh(6\beta J_2 M)}
\end{equation} 
and of the gap, $\Delta(T)$ 
\begin{equation} 
\begin{split}
\Delta=\Delta_0 &- \frac{2\sigma
b}{\kappa} \\
& -\frac{b^2}{2\kappa}\frac{\cosh(2\beta J_2 M)+9
e^{-\beta\Delta}\cosh(6\beta J_2 M)}{\cosh(2\beta J_2
M)+e^{-\beta\Delta}\cosh(6\beta J_2 M)}.
\end{split}
\end{equation}
These equations are obtained by minimizing the free energy with
respect to $h_\alpha$, $h_\beta$ and $\eta$.  The solution of
these equations is carried out using experimental values for
$\sigma$ and $\kappa$, as obtained from neutron scattering
measurements\cite{Sakurai} of the phonon spectra, which yields
$\kappa=600\Delta_0$.
\begin{figure}
      \psfrag{x3}{T$_N^b$}
      \psfrag{x1}[c]{T/T$_N^s$}
      \psfrag{x2}[c]{Intensity (arb. units)}
     \includegraphics*[width=3in]{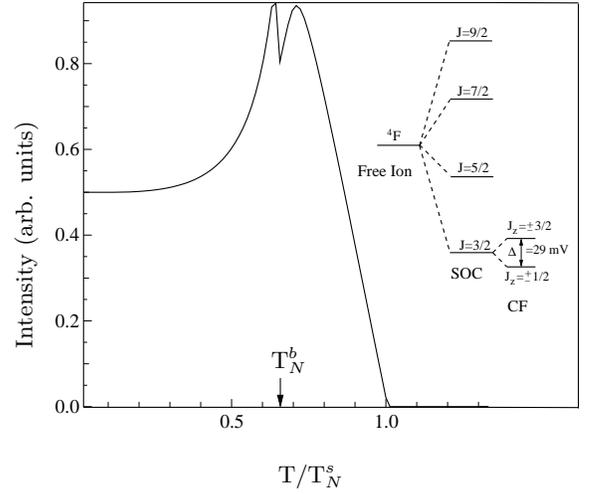}
\caption{\label{theory-a}Temperature-dependence of the half order peak
intensity from a variational mean-field calculation.}
\end{figure}

The temperature-dependence of the half-order peak intensity
$I_{(1/2,0)}(T)$ is determined by substituting the computed $M(T)$
into Eq.~(\ref{subl-mag}), with $W=0$.  Fig.~\ref{theory-a} shows
$I_{(1/2,0)}(T)$ where both the anomalous enhancement and the
depression at $T_N^b$ are manifested.  It corresponds to
$J_2/\Delta_0=0.25$ and $b=8.8\Delta_0$, comparable to reported
experimental values for CoO\cite{Herrmann,Kanamori_2}.

\begin{figure}[tb]
\begin{center}
\psfrag{x1}[c]{$T/T_N^s$}
\psfrag{x2}[c]{$I_M(T)$ (arb. units)}
\psfrag{x3}[c]{$n_e$}
\includegraphics*[width=2.75in]{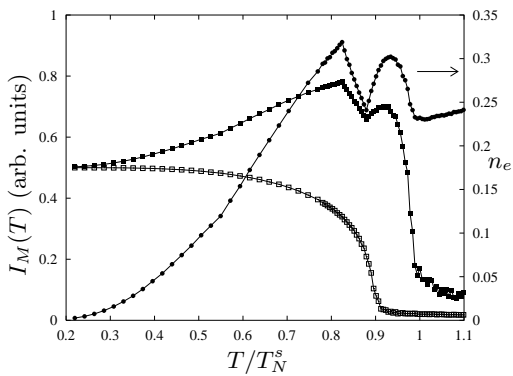}
\end{center}
\caption{\label{mc}Computed surface magnetization (filled squares), bulk
magnetization (open squares) and the excited state population (filled
circles) as a function of temperature.} 
\end{figure}

A Monte-Carlo simulation, based on the following Hamiltonian 
\begin{equation}
\begin{split}
H &= H_s + H_{sb} + H_b \\
H_a &= J_a \sum_{<nnn>} {\vec S_i^a
\cdot \vec S_j^a} + \sum_i \left({\Delta_0^s \over 2} + b_a
\sum_k\eta_{i,k}^a\right)\ {(S_{iz}^a)^2} \\ 
& \quad + {\kappa_a\over2}
\sum_i\sum_k \left(\eta_{i,k}^a\right)^2, \\ 
H_{sb} &= J_{sb}
\sum_{<nnn>} {\vec S_{i}^s \cdot \vec S_{j}^b},
\end{split}
\end{equation}
also confirms the anomalous effects.
Here, the indices $s$ and $b$ refer to surface and bulk, respectively,
and $a=s\ {\rm or}\ b$. The sum over $k$ runs over all the
nearest-neighbor O$^{-2}$ ions for a Co$^{+2}$ ion at site $i$.  We
show typical results in Fig.~\ref{mc} for a $20\times20\times33$ slab
with periodic boundary conditions. Filled squares represent the
magnetic diffraction peak intensities for the surface, $I_M^s(T)$ with
the Debye-Waller factor $W = 0$, while the open squares show the
corresponding intensities for the bulk, $I_M^b(T)$. $I_M^s(T)$
manifests both the anomalous enhancement and the dip at $T_N^b$,
similar to both experimental and mean field results. It also shows
that $T_N^s > T_N^b$.

The filled circles in Fig. 5 depict the population of the excited
state $n_e$, which mimics the behavior of $I_M^s(T)$ in the anomalous
region.  $n_e$ is appreciably higher than would be expected from
thermal activation across the gap $\Delta_0$. We also note that the
bulk spins do not populate the $j=3/2$ manifold, since the energy gap
is much higher than $T_N^b$.  

Despite the fact that both mean field and Monte-Carlo results show a
suppression in the intensity at $T_N^b$, a much more pronounced
attenuation is observed in the experiment. This difference in the
magnitude of attenuation can be attributed to the fact that the model
we used is a simple one with the simplest choice for crystal field
potential and magnetoelastic coupling. But it should be emphasized
that even with the current model, the most important features of the
experimental results, namely the enhancement and the suppresion of the
intensity at $T_N^b$ are captured.

In conclusion, the following picture emerges from the results
presented above. At low enough temperatures, the ordered AFM bulk
spins pin the surface spins to the $s_z=1/2$ state. As the
temperature increases, the bulk spin ordering decreases and the
surface spins, less constrained by the bulk, re-orient by populating
the first $s_z=3/2$ excited state.  The reorientation arises because
of two contributions to the free energy: The energy of the 3/2-state
is lowered by an amount $\sim S^2J_s$ through coupling to surface
neighbors, with $S=\pm {3\over 2}$, and the entropy is increased by
populating new states. The depression in half-order peak intensity at
$T_N^b$ is due to magnetoelastic coupling to the bulk.

This work is supported by the U.S. Department of Energy under
Grant No. DE-FG02-85ER45222.


\end{document}